# Photodiodes based in $La_{0.7}Sr_{0.3}MnO_3$ /single layer $MoS_2$ hybrid vertical heterostructures

Yue Niu[1,2], Riccardo Frisenda[1], Simon A. Svatek[1,5], Gloria Orfila[3,4,7], Fernando Gallego[6,7], Patricia Gant[1], Nicolás Agraït[1,5], Carlos León[3,4,7], Alberto Rivera-Calzada[3,4,7], David Perez De Lara[1], Jacobo Santamaria[3,4,7,*] Andres Castellanos-Gomez[6,*]

[1] Instituto Madrileño de Estudios Avanzados en Nanociencia (IMDEA-Nanociencia), Faraday 9, Ciudad Universitaria de Cantoblanco, 28049 Madrid, Spain.

[2] National Key Laboratory of Science and Technology on Advanced Composites in Special Environments, Harbin Institute of Technology, Harbin, China

[3] GFMC, Departamento de Fisica de Materiales. Universidad Complutense de Madrid, 28040 Madrid, Spain.

[4] GFMC, Instituto de Magnetismo Aplicado "Salvador Velayos", Universidad Complutense de Madrid, 28040 Madrid, Spain.

[5] Departamento de Física de la Materia Condensada and IFIMAC. Universidad Autónoma de Madrid, Madrid, E-28049, Spain.

[6] Instituto de Ciencia de Materiales de Madrid (ICMM-CSIC), Sor Juana Ines de la Cruz, 3,Cantoblanco, 28049 Madrid, Spain.

[7] Unidad Asociada "Laboratorio de heteroestructuras con aplicacion en espintronica", UCM/CSIC, Sor Juana Ines de la Cruz, 3,Cantoblanco, 28049 Madrid, Spain

E-mail:Jacobo Santamaría: jacsan@fis.ucm.es .Andres Castellanos-Gomez : andres.castellanos@csic.es

## ABSTRACT

The fabrication of artificial materials by stacking of individual two-dimensional (2D) materials is amongst one of the most promising research avenues in the field of 2D materials. Moreover, this strategy to fabricate new man-made materials can be further extended by fabricating hybrid stacks between 2D materials and other functional materials with different dimensionality making the potential number of combinations almost infinite. Among all these possible combinations, mixing 2D materials with transition metal oxides can result especially useful because of the large amount of interesting physical phenomena displayed separately by these two material families. We present a hybrid device based on the stacking of a single layer $MoS_2$ onto a lanthanum strontium manganite ($La_{0.7}Sr_{0.3}MnO_3$) thin film, creating an atomically thin device. It shows a rectifying electrical transport with a ratio of $10^3$, and a photovoltaic effect with $V_{oc}$ up to 0.4 V. The photodiode behaviour arises as a consequence of the different doping character of these two materials. This result paves the way towards combining the efforts of these two large materials science communities.

Since the first experimental isolation of graphene in 2014 [1], the interest on other layered 2D materials has kept growing [2–4]. Moreover, the community working on 2D materials is rapidly moving from the fundamental study of these atomically thin materials towards integrating them with other advanced materials to create hybrid devices. The case where an artificial layered material is created by stacking two different individual 2D layers results highly appealing [5–10], but also a great deal of experimental efforts have been paid to fabricate hybrid devices by combining 2D materials with other functional materials [11]. Lopez-Sanchez et al. and Gehring et al., for instance, fabricated artificial heterostructures by stacking 2D semiconductors on top of conventional 3D semiconductors [12–14], Velez et al. combined $MoS_2$ with organic semiconductors [15] and Jariwala et al. fabricated hybrid devices by combining mechanically exfoliated $MoS_2$ flakes with a carbon nanotube film [16].

Transition Metal Oxides (TMO), sometimes called simply complex oxides, constitute a very interesting family of materials because they display very different physical phenomena like superconductivity, ferroelectricity, ferro and antiferromagnetism, etc. The strong correlation between the several degrees of freedom of TMO (charge, spin, strain, doping) produces a rich variety of phases which result in complex phase diagrams [17,18]. Manganites are one of the paradigmatic examples of TMO showing a characteristic magnetoresistance behavior. Therefore, the combination of 2D materials with TMO is very prospective because of the rich physical phenomena that both families present separately, but the amounts of works on 2D materials/TMO heterostructures are still very scarce.

In this work we fabricate a vertical heterostructure by transferring a single-layer $MoS_2$ flake on top of a $La_{0.7}Sr_{0.3}MnO_3$ (LSMO) thin film, epitaxially grown on a $SrTiO_3$ (STO) substrate. We study the electrical transport properties of the heterostructure that behaves as a diode because of the junction between the LSMO film and $MoS_2$. This junction presents an electrical rectifying behavior (with a rectification ratio, R.R., of ~$10^3$) and upon illumination it shows photovoltaic effect. These results constitute the first steps towards combining these two large families of functional materials, opening the door for many possible 2D materials/complex oxide hybrid devices.

The LSMO film was grown by high pressure pure oxygen (3.2 mbar) RF sputtering at 900 °C on (001) oriented optically polished $SrTiO_3$ substrates. The LSMO layer was fully epitaxial and uniformly strained and had a ferromagnetic ground state. Details about sample properties and growth can be found elsewhere [19]. For the device fabrication, the LSMO film was patterned by electron beam lithography to form mesas. Firstly, MaN-2403 negative resist was spin coated on the sample, and the pattern was exposed to the electron beam. After the development of the resist, the material was etched resulting in the bottom LSMO nanostructured layer. The bottom (LSMO) electrical contact was defined evaporating Ag through a mechanical mask. Silver electrodes were checked to provide ohmic contacts to the LSMO film in agreement with previous results [20].

The structural quality of the sample was analyzed by X-ray diffraction (XRD) obtaining the pattern shown in the top inset to Figure 1. A zoom to the 001 reflection of the STO substrate is presented, showing the 001 peak of the film, and many lateral fringes at both sides, indicating the high crystalline quality of the LSMO film. In the main panel of Figure 1, the X-ray reflectivity (XRR) of the LSMO layer is plotted, presenting again many oscillations up to over 5 degrees in 2θ, as a consequence of the flatness of the film. The spacing of the maxima is a function of the thickness of the film, yielding a value of 28 ± 1 nm. The lower inset in Figure 1 presents the 3D AFM image of the etched LSMO mesa on which the $MoS_2$ will be deposited, showing the sharp edges of the LSMO nanostructure and confirming the thickness of the layer (28 ± 1 nm).

The single layer $MoS_2$ flake was deposited onto a polydimethylsiloxane (PDMS) stamp (Gelfilm by Gelpak) by mechanical exfoliation of a bulk molybdenite natural crystal (Holly Mill mine, Canada) with a SPV 224 Nitto tape. The single layer region is identified at glance by its optical contrast in transmission mode microscopy and the number of layers is double checked by quantitative analysis of the transmission mode images and by determining the energy of the A, B and C excitons from micro-reflectance spectroscopy [21,22]. Figure 2(a) shows a transmission mode optical image of the $MoS_2$ flake onto the PDMS stamp prior to transfer (the inset shows the transmission mode image of the same flake transferred onto the LSMO film). Figure 2(b) shows the micro-reflectance spectra acquired on the $MoS_2$ flake before being transferred to the LSMO surface, used to determine the number of $MoS_2$ layers. Then the flake was transferred onto the LSMO surface by an all-dry deterministic transfer technique, described in detail in Ref. [23].

The electrical transport properties of the $MoS_2$/LSMO stack are characterized with a homebuilt probe station equipped with carbon fiber microprobes that can be accurately placed onto mechanically exfoliated flakes (the diameter of the fiber microprobes is of only 7 μm) without damaging them. Their Young modulus is ~ 280 GPa and the spring constant of a typical carbon fiber probe (~1 mm long) is ~0.02N/m,[24,25] allowing for very gentle mechanical contact without damaging the flakes. The electrical contact between a carbon fiber probe and $MoS_2$ flakes has been reported to be Ohmic [26]. More details on this experimental setup and measurements can be found in Ref. [26]. Figure 2(c) is a sketch of the experimental configuration employed to probe the electrical properties of the stack. A source measure unit (Keithley 2450) is used to perform the current-voltage measurements. The light source is provided by a high-power LED source (455 nm of illumination wavelength with a power density up to 0.64 $W/cm^2$).

Figure 3 shows the current vs. bias voltage trace (*IV* hereafter) measured in the dark state, showing a marked rectifying behavior. The inset in Figure 3 displays the same *IV* in logarithmic scale (with the current in absolute value) to facilitate the estimation of the rectification ratio of the diode. The device, despite of being atomically thin, presents a very large rectification ratio of $\sim 10^3$. Note that Ag and LSMO have very similar work functions (4.7 eV and 4.8 eV respectively), and thus Ag acts as a good ohmic contact to LSMO [19]. On the contrary, the electron affinity of $MoS_2$ is lower (4.2 eV). Moreover, the Fermi level in LSMO is close to valence band and in $MoS_2$ the Fermi level is quite close to the conduction band. Therefore, the LSMO behavior can be described as a hole metal (P++), and the $MoS_2$ a N+ semiconductor. It is therefore expected that electrons are transferred in the junction from the $MoS_2$ into the LSMO, eventually leading to a decrease in the concentration of electrons in $MoS_2$. Similar charge transfer has been observed in other metal oxide - transition metal dichalcogenide interfaces [27,28]. Nevertheless, our results indicate that in the $MoS_2$/ LSMO junctions the $MoS_2$ behaves as the N side and LSMO as the P side of a diode. In general, in atomically thin PN-junctions there is no built-in field and thus the carrier transport occurs by tunneling processes [5]. However, in this case the thickness of the LSMO film allows carrier diffusion and drift processes. Therefore, a built-in field may be generated in the LSMO side to level the chemical potentials. Notice the reversibility of *IV* curves measured sequentially in the dark (Supporting Information, Figure S6 and S7) excludes oxygen vacancy generation in the range of applied electric fields of the experiment, which would show up in the *IV* curves as resistive switching effects [29]. Assuming that the voltage corresponding to the difference between the work functions of both materials drops in a space charge region in the LSMO of a width of the Thomas Fermi screening length (typically between 1 and 2 nm), an electric field can be estimated in the range $2\text{-}5\times10^6$ $V\cdot cm^{-1}$. This field is screened by a sheet carrier density of the order $10^{12}$ $cm^{-2}$ quite close to the typical values of the sheet carrier density of $MoS_2$ [30].

After characterizing diode-like electrical transport properties of the stacked device we study the performance of the device upon illumination. We project a high-power LED source, forming a spot of 400 µm in diameter, onto the stacked $MoS_2$/LSMO. Figure 4(a) shows the *IV* characteristics of the device in dark and upon illumination with increasingly high power density. The forward current increases monotonically with the illumination power density because of the photogeneration of electron-hole pairs that are separated by the applied drain-source bias voltage (photoconductive generation mechanism). At zero applied bias, the built-in electric field in the interface (due to the different doping character of the two materials) separates the photogenerated electron-hole pairs producing a finite current (so called $I_{sc}$ short circuit current), which is a fingerprint of the photovoltaic effect. We ruled out the photogeneration of current at the Ag/LSMO and carbon fibre/$MoS_2$ interfaces as they present Ohmic contact behavior.[19,26] Although scanning photocurrent mapping could be an independent way to proof the negligible effect of the Ag/LSMO and carbon fibre/$MoS_2$ interfaces in the photocurrent generation but its implementation in combination with the carbon fiber top electrode results technically challenging and it lays beyond the scope of this manuscript. The voltage (at zero current) generated in the junction upon illumination (open circuit voltage, $V_{oc}$) and it is another characteristic of the photovoltaic effect. Figure 4(b) shows a zoom in the *IV* characteristics of Figure 4(a) to facilitate the identification of the short circuit current and open circuit voltage generated upon illumination. Figure 4(c) shows the optical power dependence of these values displaying a linear dependence for $I_{sc}$ and logarithmic dependence for $V_{oc}$, characteristics of photodiodes. The $V_{oc}$ saturates at values close to 0.4 V even without optimizing the carrier extraction electrodes. The comparison with other heterojunctions in the literature shows that the $V_{oc}$ value of our complex oxide/2D semiconductor is comparable to that of other devices based on vertical stacking of 2D semiconductors (see Table 1) and also the filling factor (FF) of our device is the largest reported so far for these kind of vertical heterostructure devices. Table 1 also compares the short circuit current density of different devices. Note that in our device the contact area is not well-defined but considering an area of 50 $\mu m^2$ (a conservative value taking into account the diameter of 7 µm of the fiber) the $J_{sc}$ value of our device is ~1 $mA/cm^2$, which is a moderate value compared with other vertical heterostructure. Nonetheless, our device also presents a unique feature: the whole stack is almost transparent (see inset in Figure 2(a)) while the rest of listed stacks are fabricated on top of opaque substrates.

In summary, we presented a hybrid device based on the artificial stacking of single-layer $MoS_2$ onto a LSMO thin film. The different doping character of these materials is exploited to build up an atomically thin photodiode device that shows a remarkably high rectification ratio (~$10^3$) and upon illumination displays photovoltaic effect with $V_{oc}$ up to 0.4 V. The results presented here constitute the first step towards combining the efforts of two large materials science communities, the two-dimensional materials and the TMO to fabricate optimized artificial materials combining suitable members of each family.

## Acknowledgements

Work at IMDEA was supported by MINECO (Ramón y Cajal 2014 program RYC-2014-01406, MAT2014-58399-JIN, FIS2015-67367-C2-1-P), the Comunidad de Madrid (MAD2D-CM Program (S2013/MIT-3007)) and NANOFRONTMAG-CM program (S2013/MIT-2850) and the European Commission under the Graphene Flagship (contract CNECTICT-604391) and FP7 ITN MOLESCO (project no. 606728). RF acknowledges support from the Netherlands Organisation for Scientific Research (NWO) through the research program Rubicon with project number 680-50-1515. YN acknowledges the grant from the China Scholarship Council (File NO. 201506120102). Work at UCM supported by Spanish MINECO through grants MAT2014-52405-C02-01 and by CAM through grant CAM S2013/MIT-2740. Work at ICMM is supported by Spanish MINECO through grant MAT2014-52405-C02-02

**Supporting Information**:

Supporting Information includes:

1) Thickness determination from differential reflectance data and quantitative analysis of the transmission mode images
2) Determination of the resistance of the carbon fiber probe
3) Electrical power generated through the photovoltaic effect
4) Measurements on another $MoS_2$/LSMO device
5) Measurements on multilayer $MoS_2$/LSMO devices
6) Current vs. voltage characteristics of a LSMO device with silver contacts

# Figures

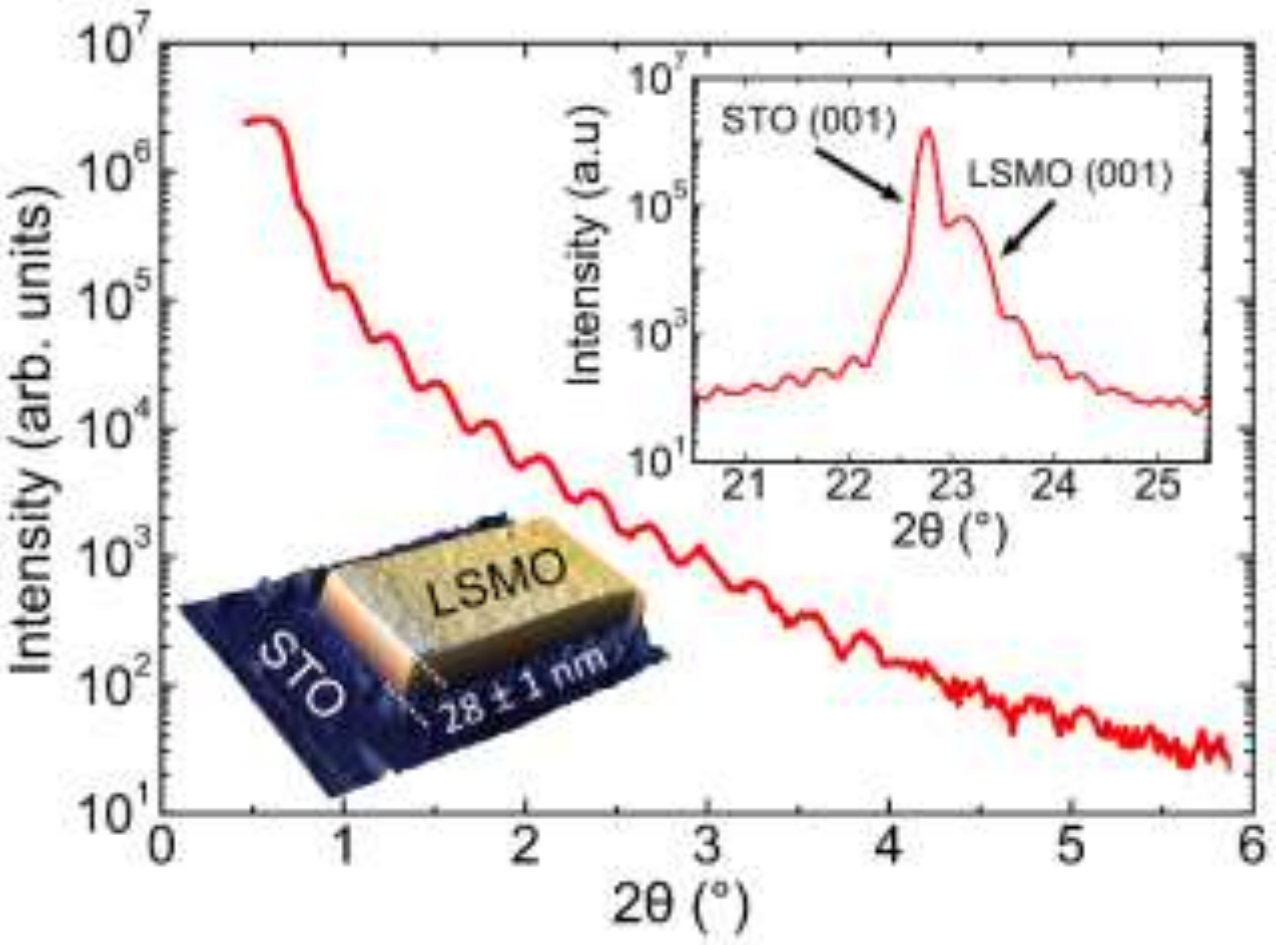

**Figure 1** X-ray diffraction pattern (top inset), and reflectivity (main) of a 28 nm LSMO film grown on STO (001) substrate. The bottom inset is an AFM 3D image of the mesa that will contain the $MoS_2$ flake.

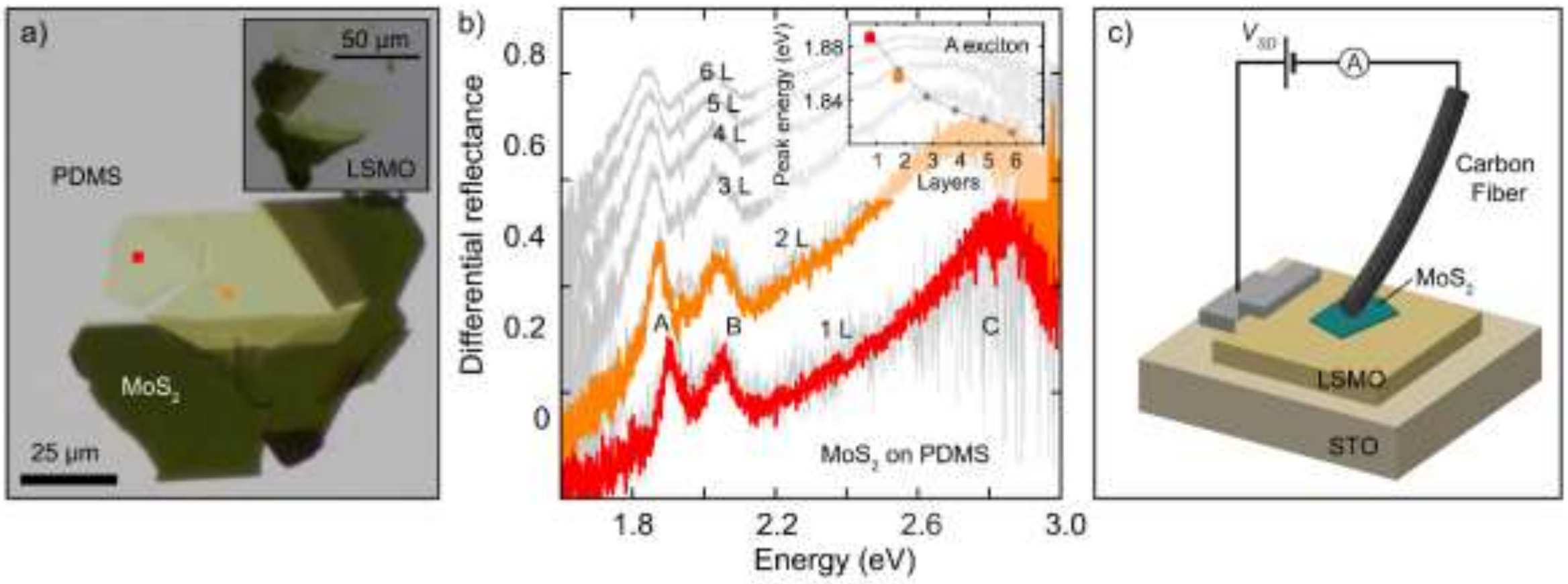

**Figure 2** (a) Transmission mode optical image of a $MoS_2$ flake on a PDMS stamp before its transfer and after being transferred onto the LSMO thin film (inset). (b) Differential reflectance spectra measured on the $MoS_2$ flake before its transfer to determine its number of layers. The spectra in red and orange have been acquired at the positions highlighted with squares of the same color in (a). The panel also includes the spectra acquired for reference $MoS_2$ samples with different number of layers (gray). The inset in (b) shows the comparison between the energy of the A exciton measured at the locations shown in (a) and on the reference samples to determine the number of layers. (c) Sketch of the experimental configuration to measure the electrical transport across the $MoS_2$/LSMO stack.

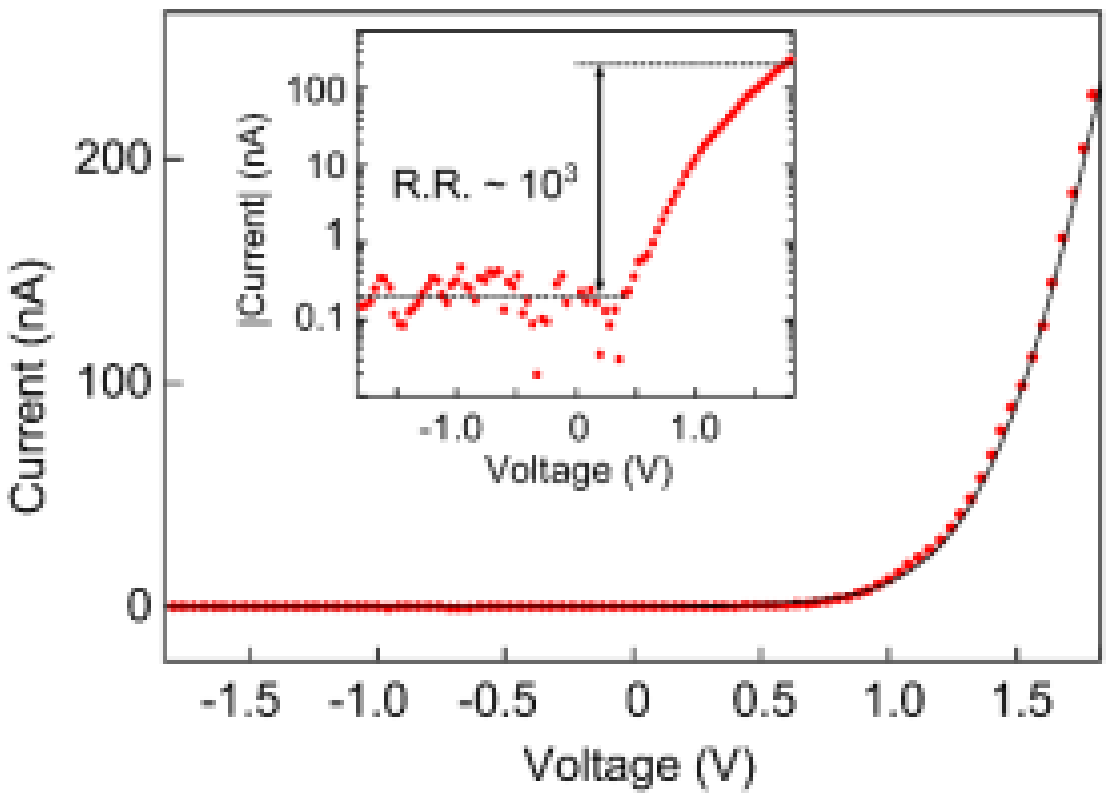

**Figure 3** Current vs. bias voltage characteristics of the single-layer $MoS_2$/LSMO device. The inset shows the same dataset in semi-logarithmic scale and displaying the absolute value of the current.

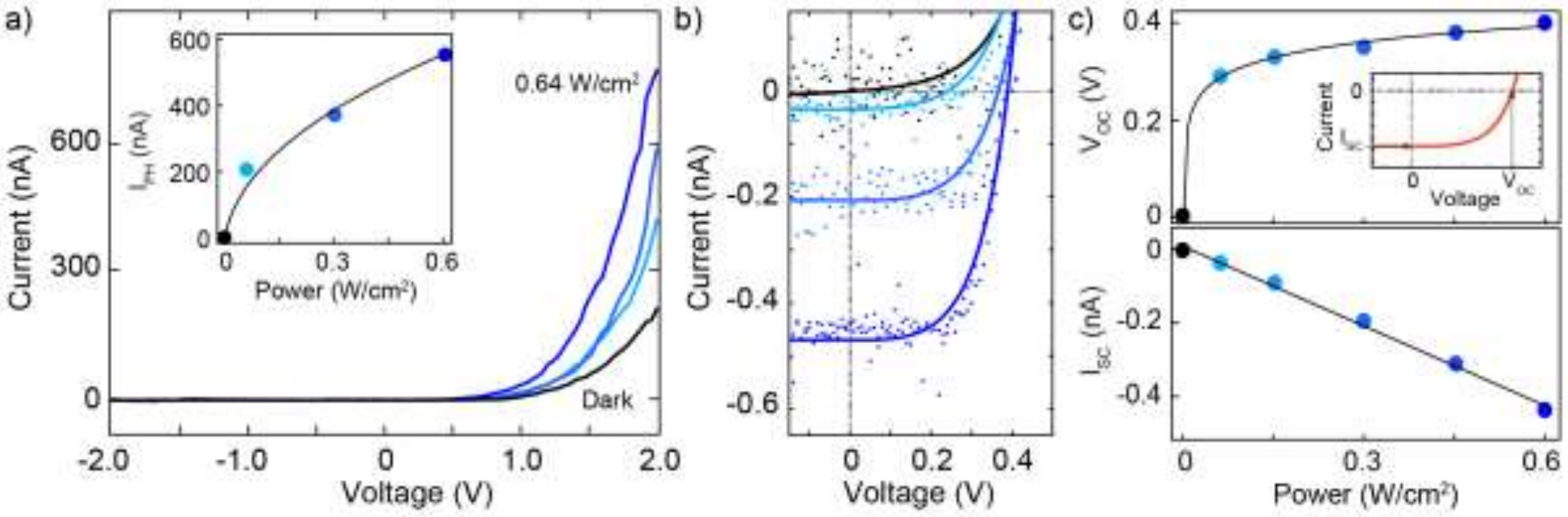

**Figure 4** (a) Current vs. bias voltage characteristics of the device in dark and upon illumination with increasingly high illumination power density. The inset shows the photogenerated current vs. the power density. (b) A zoomed in plot around zero bias to show the photovoltaic effect upon illumination. (c) Power dependence of the open circuit voltage and the short circuit current. Illumination wavelength 455 nm.

| Vertical stack | | $V_{oc}$ (Volts) | $J_{sc}$ (mA/cm$^2$) | F.F(%) | Reference |
|---|---|---|---|---|---|
| 2D/3D semiconductor | 1L-$MoS_2$/p-Si | 0.41 | 22.4 | 0.57 | 31 |
| | 1L-$MoS_2$/p-Si | 0.55 | 100 | 0.1 | 13 |
| | FL-BP/GaAs | 0.55 | 10000 | 0.30 | 14 |
| | FL-$Bi_2Te_3$/p-Si | 0.24 | 0.96 | 0.38 | 32 |
| 2D/complex oxide | 1L-$MoS_2$/28 nm LSMO | 0.40 | 1.0 | 0.69 | This work |
| 2D/organic semiconductor | 2L-$MoS_2$/Cu-phthalocyanine | 0.60 | 0.25 | 0.26 | 15 |
| 2D/2D | 1L-$MoS_2$/1L $WSe_2$ | 0.45 | 0.85 | 0.29 | 5 |
| | 1L-$MoS_2$/1L $WSe_2$ | 0.55 | 0.52 | 0.49 | 33 |
| | FL-$MoS_2$/FL-$WSe_2$ | 0.27 | 1000 | 0.44 | 34 |
| | 1L-$MoS_2$/FL-BP | 0.30 | 9.5 | 0.32 | 8 |

**Table 1.** Comparison between the open circuit voltages reported for vertical PN junction devices based on the stacking of semiconducting 2D materials.

# Supporting Information:

# Photodiodes based in $La_{0.7}Sr_{0.3}MnO_3$ /single layer $MoS_2$ hybrid vertical heterostructures

Yue Niu[1,2], Riccardo Frisenda[1], Simon A. Svatek[1,5], Gloria Orfila[3,4,7], Fernando Gallego[6,7], Patricia Gant[1], Nicolás Agraït[1,5], Carlos León[3,4,7], Alberto Rivera-Calzada[3,4,7], David Perez De Lara[1], Jacobo Santamaria[3,4,7,*] Andres Castellanos-Gomez[6,*]

[1] Instituto Madrileño de Estudios Avanzados en Nanociencia (IMDEA-Nanociencia), Faraday 9, Ciudad Universitaria de Cantoblanco, 28049 Madrid, Spain.

[2] National Key Laboratory of Science and Technology on Advanced Composites in Special Environments, Harbin Institute of Technology, Harbin, China

[3] GFMC, Departamento de Fisica de Materiales. Universidad Complutense de Madrid, 28040 Madrid, Spain.

[4] GFMC, Instituto de Magnetismo Aplicado "Salvador Velayos", Universidad Complutense de Madrid, 28040 Madrid, Spain.

[5] Departamento de Física de la Materia Condensada and IFIMAC. Universidad Autónoma de Madrid, Madrid, E-28049, Spain.

[6] Instituto de Ciencia de Materiales de Madrid (ICMM-CSIC), Sor Juana Ines de la Cruz, 3,Cantoblanco, 28049 Madrid, Spain.

[7] Unidad Asociada "Laboratorio de heteroestructuras con aplicacion en espintronica", UCM/CSIC, Sor Juana Ines de la Cruz, 3,Cantoblanco, 28049 Madrid, Spain

E-mail:Jacobo Santamaría: jacsan@fis.ucm.es .Andres Castellanos-Gomez : andres.castellanos@csic.es

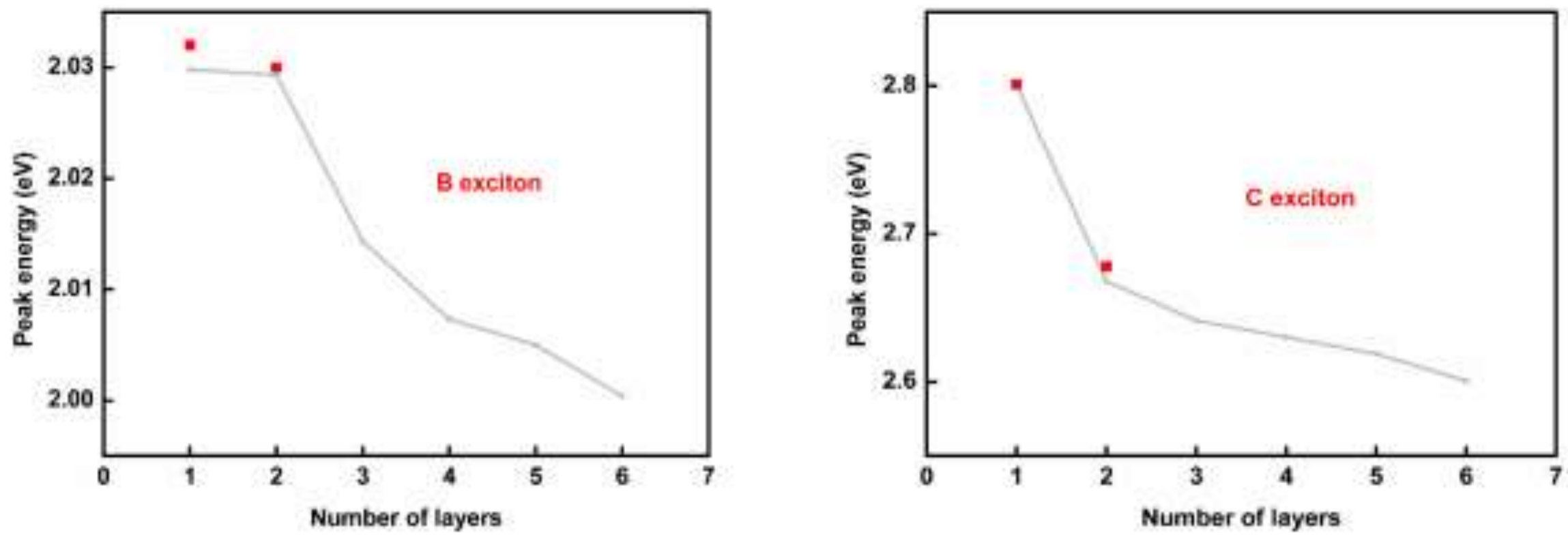

**Figure S1** Thickness dependence of the B (left panel) and C (right panel) exciton peaks extracted from the spectra in Figure 2 of the main text.

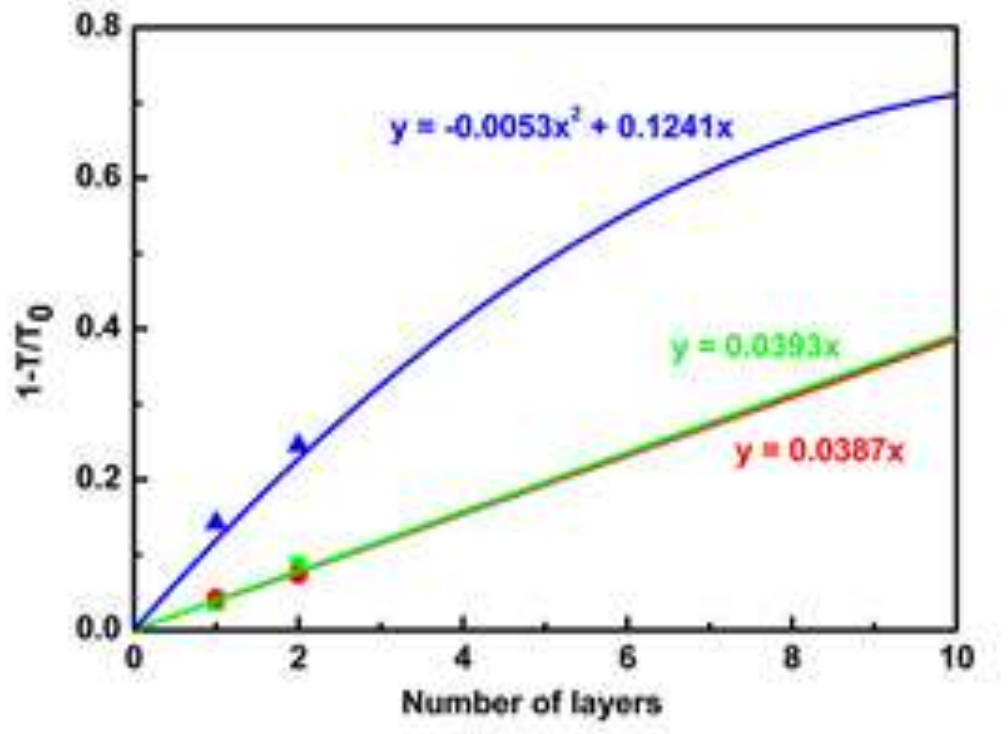

**Figure S2** Thickness determination from the quantitative analysis of the transmission mode optical image of the flake on PDMS (Figure 2a of the main text). The red, green and blue lines and symbols correspond to the red, green and blue channels of the digital camera. The symbols are the experimental absorption data measured at the two locations highlighted in Figure 2a. The solid lines show the thickness expected absorption of $MoS_2$.

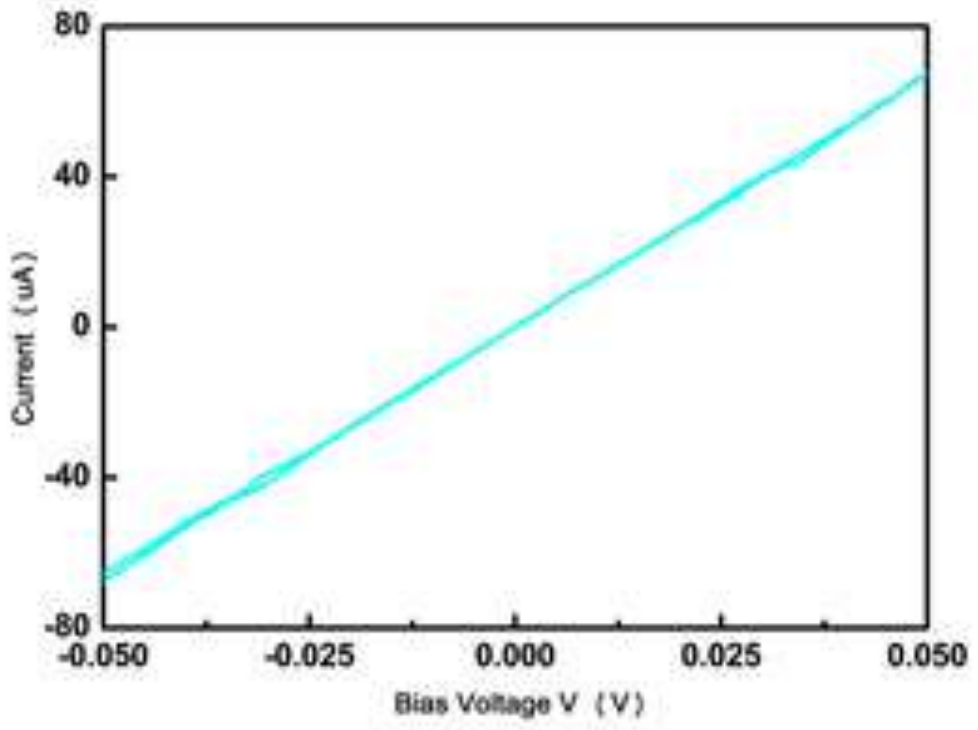

**Figure S3** Current vs. bias voltage characteristics of the carbon fiber probe in contact with a gold surface to determine the resistance of the fiber probe, R = 750 Ω.

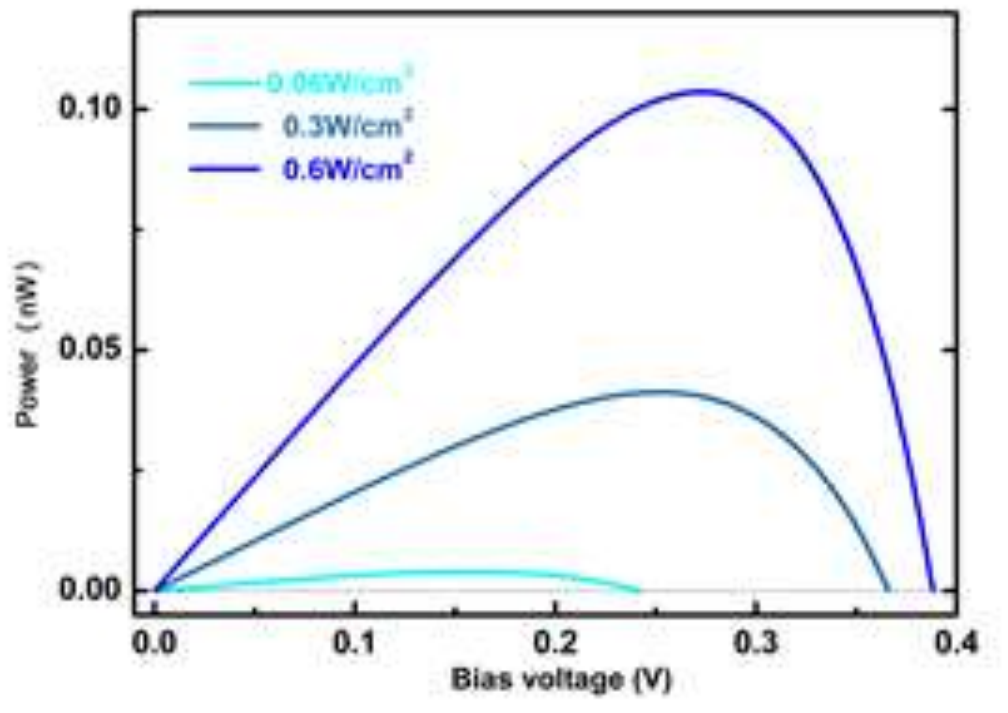

**Figure S4** Photovoltaic generated power, calculated from the current vs. bias voltage characteristics shown in Figure 4b of the main text.

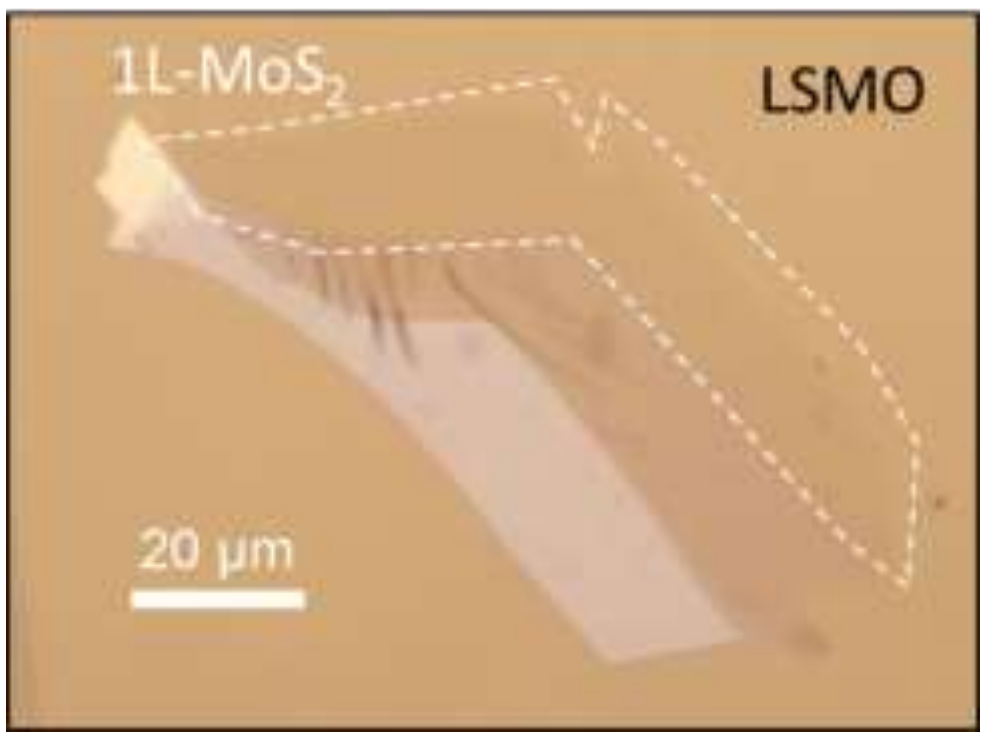

**Figure S5** Epi-illumination optical image of another single-layer $MoS_2$ flake transferred onto a LSMO thin film.

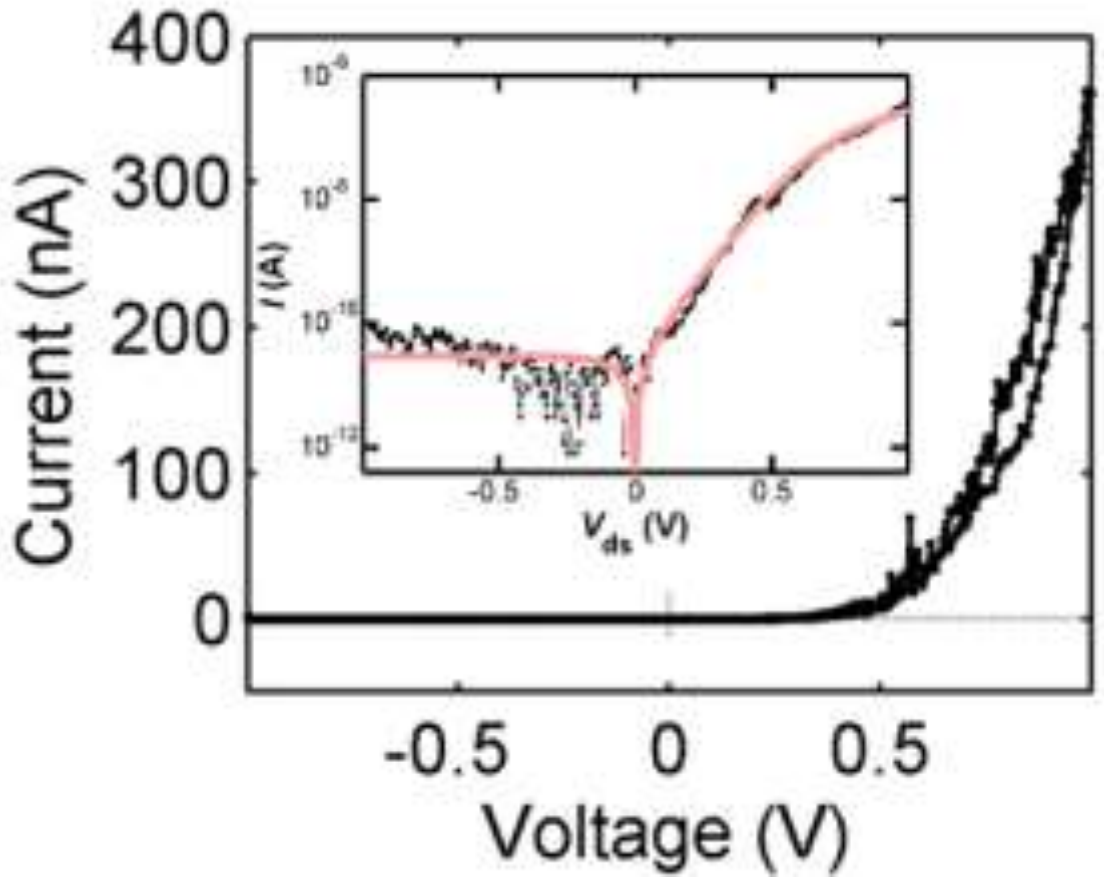

**Figure S6** Current vs. bias voltage characteristics of the device shown in Figure S5. The inset shows the same dataset in semi-logarithmic scale.

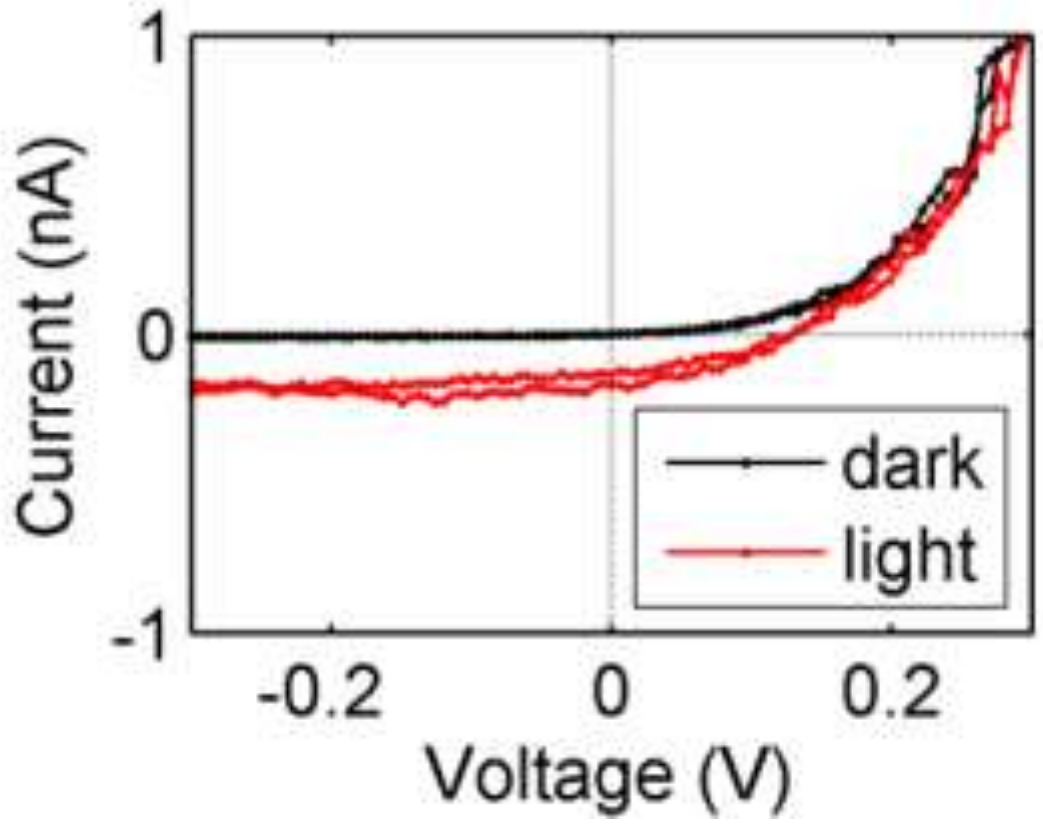

**Figure S7** Current vs. bias voltage characteristics of the device shown in Figure S5 in dark and upon illumination. The characteristics upon illumination shows photovoltaic effect with $V_{oc}$ = 0.12 V and $I_{sc}$ = 0.2 nA. Illumination: 455 nm of wavelength and power density of 235 mW/cm$^2$.

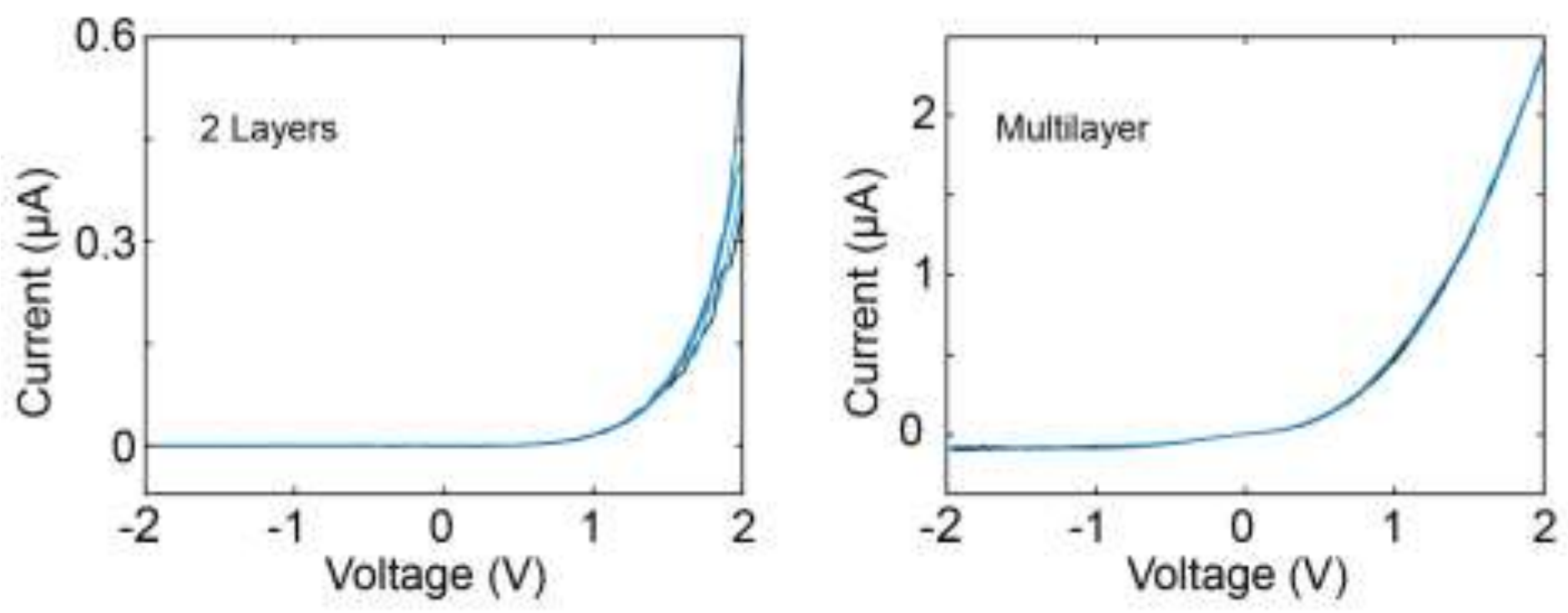

**Figure S8** Current *vs*. bias voltage characteristics of a bilayer (left) and a multilayer (right) from device 1.

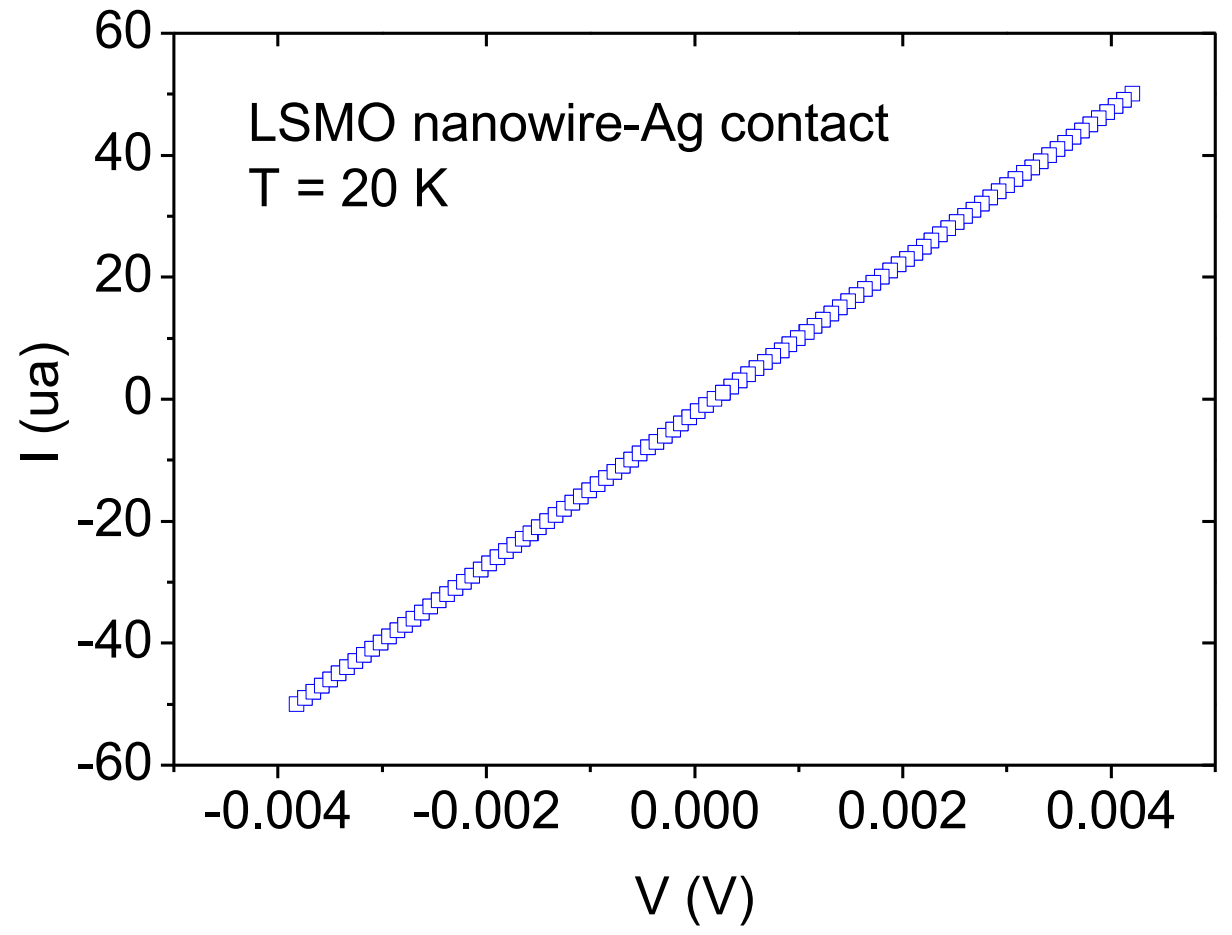

**Figure S9**. Current *vs*. bias voltage characteristics measured in a LSMO nanowire contacted with silver electrodes showing a very high current density with linear IV, indicative of Ohmic contact.